# Priority based Bandwidth Adaptation for Multi-class Traffic in Wireless Networks

Mostafa Zaman Chowdhury[1], Yeong Min Jang[1], and Zygmunt J. Haas[2]
[1]*Department of Electronics Engineering, Kookmin University, Seoul, Korea.*
[2]*Wireless Networks Lab, Cornell University, Ithaca, NY, 14853, U.S.A*
*E-mail: mzceee@yahoo.com, yjang@kookmin.ac.kr, zhaas@cornell.edu*

***Abstract***

*The bandwidth adaptation is the technique that allows the flexibility in bandwidth allocation for a call. Using the bandwidth adaptation technique, the number of call admission in the system can be increased significantly. In this paper we propose a priority based bandwidth adaptation scheme that can release multi-level of bandwidth from the existing calls to accept call requests. The amount of released bandwidth is based on the number of existing bandwidth adaptive calls and the priority of requesting traffic call. This priority scheme does not reduce the bandwidth utilization. Moreover, the proposed bandwidth adaptation strategy provides significantly reduced call blocking probability for the higher priority traffic calls. The performance analyses show the improvement of the proposed scheme.*

## 1. Introduction

The trend of the wireless communication is the increase of various types of high data rate traffic. However, the wireless capacity is not increasing as required to fulfill the demands. The bandwidth adaptation is one of the well-known technique that allows the flexibility in bandwidth allocation for the calls. This technique provides significantly increased number of call admission in the system. The quality of service (QoS) adaptability e.g., bandwidth adaptation has been used by several schemes (e.g., [1], [2]]) to reduce the call blocking probability. The adaptive QoS schemes [3], [4] proved more flexible and efficient in guaranteeing QoS than the guard channel schemes [1]. However, the adaptive bandwidth allocation schemes enlarge the call duration for some of bandwidth adaptive calls. In this paper, we present a priority based bandwidth adaptation scheme that allows reclaiming some of the allocated bandwidth from already admitted bandwidth adaptive traffic calls, as to accept higher priority calls, when the system's resources are running low. Therefore, our proposed scheme can reduce the overall forced call termination probability significantly as well as provides reduced call blocking probability for the higher priority users. The proposed scheme reserves some releasable bandwidth to accept higher priority calls by providing priority based multi-level bandwidth adaptation. The proposed scheme only increases the call blocking probability of lower priority traffic calls compared to the non-priority bandwidth adaptive scheme.

The rest of this paper is organized as follows. Section 2 presents our proposed multi-level bandwidth allocation scheme. Performance analyses results of the proposed scheme are presented and compared with other schemes in Section 3. Finally, Section 4 concludes our work.

## 2. Priority based Bandwidth Adaptation

The QoS of non-real-time traffic calls are normally adaptive [5]. Therefore, normally they can release some of of their occupied resources to increase the number of call admission in the system. The limit of bandwidth adaptation of an existing traffic call in our scheme is based on the priority of the requested call. The existing calls can make higher amount of empty bandwidth to accept a higher priority traffic call compared to the lower priority calls. The main goal of the proposed scheme is to maximize the number of call admission in the system by providing bandwidth adaption technique as well as to offer significantly reduced call blocking probability for the higher priority calls. The priority is given by providing different level of bandwidth adaption to accept different types of call requests. The basic idea for our proposed scheme is shown in Fig. 1. A call of *m-th* class traffic can be allocated by different level of bandwidth. The proposed bandwidth allocation scheme is characterized by few bandwidth degradation [1]–[4], [6] factors $\gamma_m$ *and* $\gamma_{m,p}$, respectively, are defined for each class $m$ traffic, as: the portion of the bandwidth that has been already degraded of an admitted call, the maximum portion of the bandwidth of an admitted call that can be degraded to accept a call request of class $p$ type traffic ($p$=0 represents the highest priority handover calls of any types of traffic and $p=M$ represents lowest priority traffic calls). Therefore, to accept *p-th* priority call, the existing *m-th* class traffic call can be degraded to the maximum limit of the $\gamma_{m,p}$ portion. If there are $M$ number of classes traffic calls, then the figure shows that the system can release maximum amount of bandwidth to accept a handover call request whereas the system can release minimum amount of bandwidth to accept a call request of *M-th* class traffic. Also, the required minimum bandwidth to accept a higher priority call is lower than that of lower priority calls.

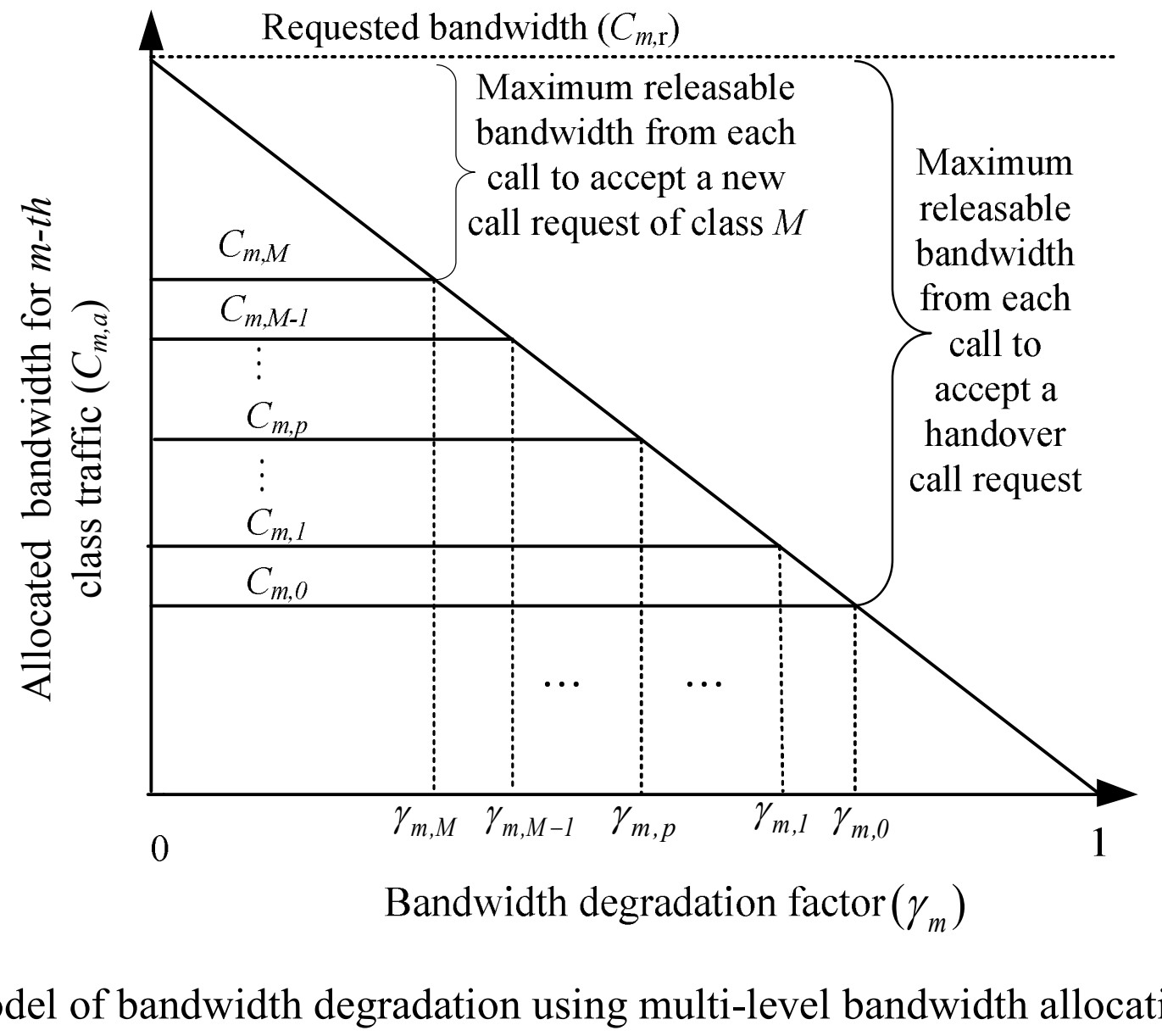

**Fig. 1.** Basic model of bandwidth degradation using multi-level bandwidth allocation for the calls of *m-th* class traffic.

The bandwidth-degradation factors relate to the bandwidth allocations as follows:

$$\gamma_m = \frac{C_{m,r} - C_{m,a}}{C_{m,r}} \quad (1)$$

$$\gamma_{m,p} = \frac{C_{m,r} - C_{m,p}}{C_{m,r}} \quad (2)$$

where $C_{m,r}$ represents the bandwidth requested by a call of the *m-th* class traffic. A call of *p-th* class ($p$=0 represents the handover calls of any types of traffic) can be accepted by the system only if the condition $C_{m,a} \geq C_{m,p}$ (for all the traffic classes of $m$=1… $M$) holds after a call of traffic class $p$ is accepted..

From the Fig. 1 it can be clearly stated that,

$$1 > \gamma_{m,0} \geq \gamma_{m,1} \geq \cdots \geq \gamma_{m,m} \geq \cdots \geq \gamma_{m,M} \geq 0 \quad (3)$$

The releasable amount of bandwidth from each of the existing *m-th* class traffic calls to accept a call of *p-th* priority traffic class is calculated as,

$$C_{release,m(p)} = C_{m,r} - C_{m,p} = \gamma_{m,p} C_{m,r} \quad (4)$$

From (1)-(4) and Fig. 1, it is clear that the system is more adaptive to accept higher priority call requests. Therefore, the proposed scheme is able to give higher priority to higher priority traffic calls in terms of call blocking probability.

## 3. Performance Analysis

Now we present the performance of our proposed scheme in this section. Table 1 shows the basic assumptions for the assumed traffic classes. The ratio of the number of requested calls (voice: web-browsing: video: background) is considered as 3:3:1:2. Considering the average call duration of 120 sec during the condition of no bandwidth degradation, the average cell dwell time is found to be 240 sec. During the bandwidth degraded condition, we considered the average call duration is state dependent i.e., more than 120 sec. We assume that the system capacity is 6 Mbps.

Fig. 2 shows one example of bandwidth release from the existing bandwidth adaptive calls. We have shown the amount of bandwidth release from each of the existing web-browsing calls with the increase of traffic congestion. Other bandwidth adaptive calls also release bandwidth with the similar manner. The figure shows that the bandwidth allocation for each of the web-browsing call is decreased with the increase of traffic load. At the light traffic condition, a web-browsing call can release bandwidth to accept any types of call request. However, for the higher traffic load condition, the system permits to release bandwidth from the existing web-browsing traffic calls only to accept higher priority traffic call requests.

**Table 1.** Basic assumptions for the analysis

| Traffic class (m) | Requested bandwidth by each call | $\gamma_{m,0}$ | $\gamma_{m,p}$ ($p=1,2,3,4$) |
|---|---|---|---|
| Conversational voice (m=1) | 32 kbps | 0 | 0 |
| Interactive web-browsing (m=2) | 120 kbps | 0.6 | $0.95\gamma_{2,p-1}$ |
| Streaming video (m=3) | 256 kbps | 0.7 | $0.95\gamma_{3,p-1}$ |
| Background (m=4) | 60 kbps | 0.8 | $0.95\gamma_{4,p-1}$ |

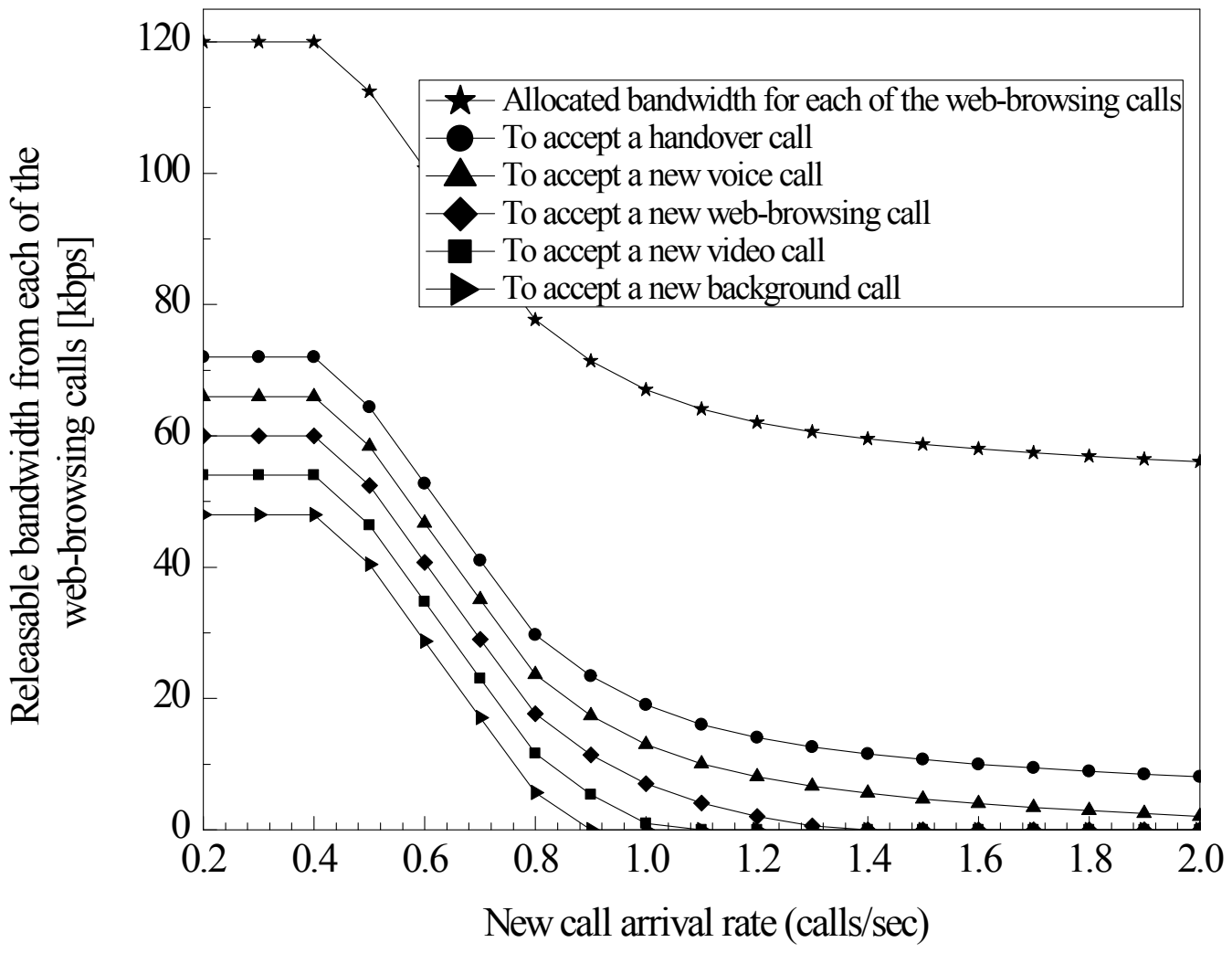

**Fig. 2.** The amount of bandwidth release from each of the existing web-browsing calls with the increase of traffic congestion.

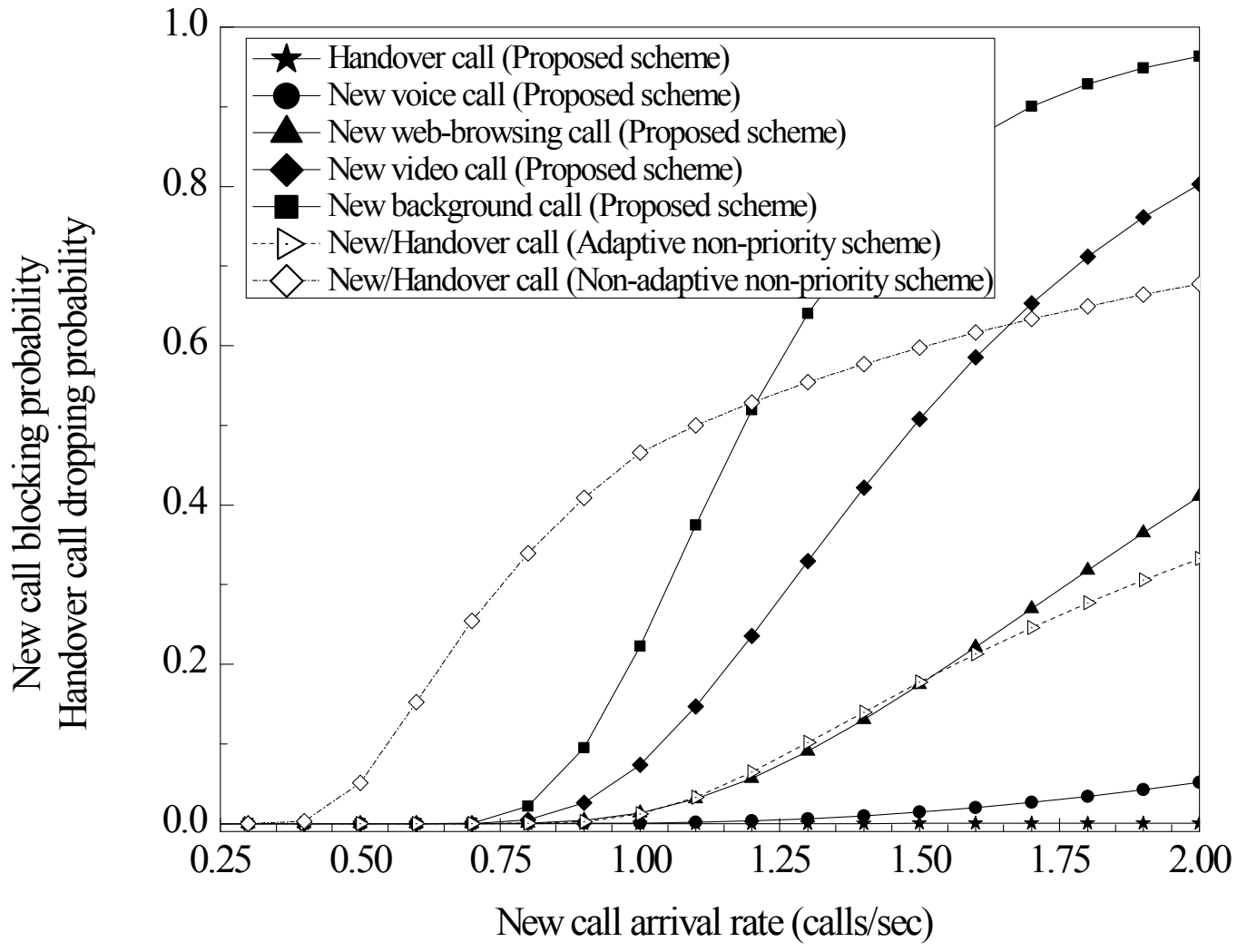

**Fig. 3.** A comparison of new call blocking probability and handover call dropping probability.

Fig. 3 shows that the proposed scheme provides negligible handover call dropping probability even for very high traffic load condition. The provision of maximum level of bandwidth adaptation without the priority of calls cannot provide acceptable handover call dropping probability and new call blocking probability of the higher priority traffic calls. Bandwidth non-adaptive non-priority scheme creates very high handover call dropping probability and also very high call blocking probability for the higher priority users due to the absence of bandwidth adaptation technique and the priority of traffic classes.

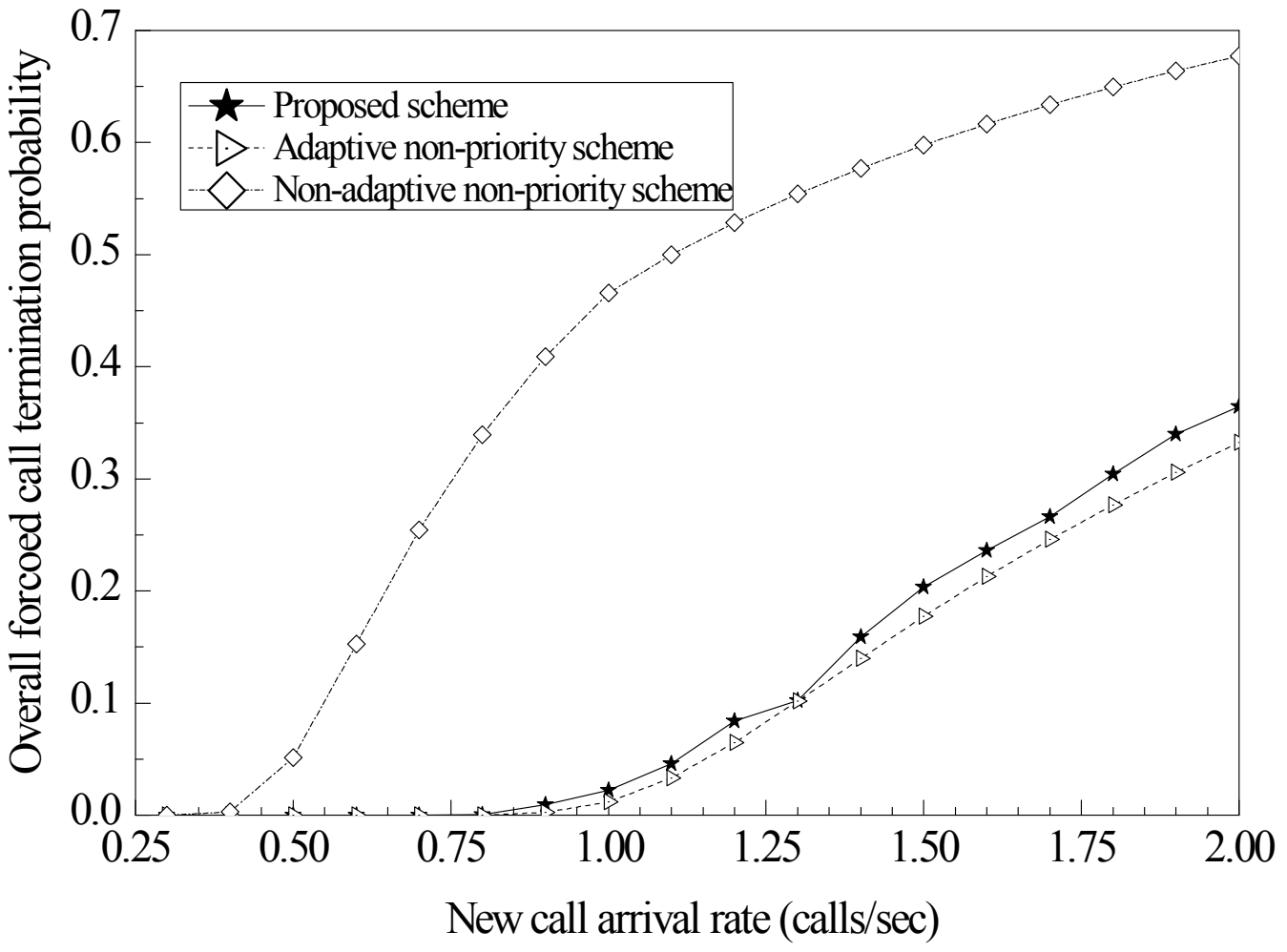

**Fig. 4.** A comparison of overall forced call termination probability.

Fig. 4 shows that our scheme provides almost equal forced call termination probability compared to the adaptive non-priority scheme. The priority of traffic classes does not increase the overall forced call termination probability significantly.

## 4. Conclusions

The bandwidth adaptation scheme is able to increase the number of call admission in the system. On the other hand, the multi-level of bandwidth adaptation technique provides the priority of traffic classes. The proposed scheme offers more bandwidth degradation of the calls to support higher priority traffic calls compared to the lower priority calls. Therefore, our proposed scheme provides priority of traffic calls as well as bandwidth adaptation. As a result, to give the priority of traffic calls, the overall forced call termination probability is not increased significantly. The proposed scheme seems to be very effective approach to make the efficient utilization of the limited wireless resources.

## Acknowledgments

This work was supported by the IT R&D program of MKE/KEIT [10035362, Development of Home Network Technology based on LED-ID].

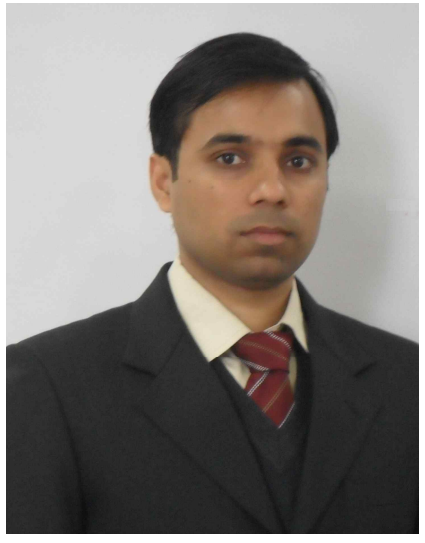

**Mostafa Zaman Chowdhury** received his B.Sc. in Electrical and Electronic Engineering from Khulna University of Engineering and Technology (KUET), Bangladesh in 2002. In 2003, he joined the Electrical and Electronic Engineering department of KUET as a faculty member. He received his M.Sc. in Electronics Engineering from Kookmin University, Korea in 2008. Currently he is working towards his Ph.D. degree in the department of Electronics Engineering at the Kookmin University, Korea. His research interests include convergence networks, QoS provisioning, mobility management, femtocell networks, and VLC networks.

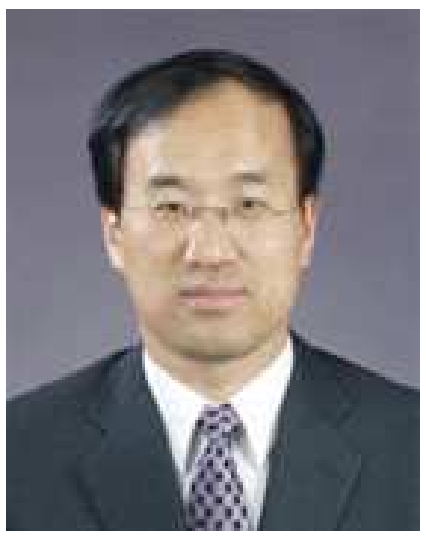

**Yeong Min Jang** received the B.E. and M.E. degree in Electronics Engineering from Kyungpook National University, Korea, in 1985 and 1987, respectively. He received the doctoral degree in Computer Science from the University of Massachusetts, USA, in 1999. He worked for ETRI between 1987 and 2000. Since September 2002, he is with the School of Electrical Engineering, Kookmin University, Seoul, Korea. His research interests include IMT-advanced, radio resource management, femtocell networks, Multi-screen convergence networks, and WPANs.

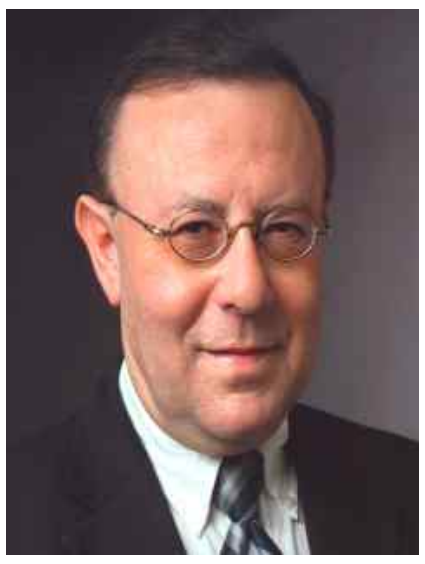

**Zygmunt J. Haas** received his B.Sc. in ECE in 1979, his M.Sc. in EE in 1985, and his Ph.D. from Stanford University in 1988. Subsequently, he joined the AT&T Bell Laboratories in the Network Research Department. There he pursued research on wireless communications, mobility management, fast protocols, optical networks, and optical switching. From September 1994 till July 1995, Dr. Haas worked for the AT&T Wireless Center of Excellence, where he investigated various aspects of wireless and mobile networking, concentrating on TCP/IP networks. In August 1995, he joined the faculty of the School of Electrical and Computer Engineering at Cornell University. He directs the *Wireless Networks Laboratory (WNL)*, an internationally recognized research group specializing in ad hoc and sensor networks. His interests include: mobile and wireless communication and networks, biologically-inspired networks, and modeling of complex systems.